\documentclass[12pt, preprint]{aastex61}
\usepackage{graphicx}
\usepackage{multirow}

\received{}
\revised{}
\accepted{}

\submitjournal{ApJ}

\shorttitle{confined flare with double HXR peaks}
\shortauthors{Ning et al.}

\begin{document}

\title{Two-stage energy release process of a confined flare with double HXR peaks}

\correspondingauthor{Yao Chen}
\email{yaochen@sdu.edu.cn}

\author{Hao Ning}
\affil{Shandong Provincial Key Laboratory of Optical Astronomy and Solar-Terrestrial Environment, and Institute of Space Sciences, Shandong University, Weihai, Shandong 264209, China}

\author{Yao Chen}
\affiliation{Shandong Provincial Key Laboratory of Optical Astronomy and Solar-Terrestrial Environment, and Institute of Space Sciences, Shandong University, Weihai, Shandong 264209, China}

\author{Zhao Wu}
\affiliation{Shandong Provincial Key Laboratory of Optical Astronomy and Solar-Terrestrial Environment, and Institute of Space Sciences, Shandong University, Weihai, Shandong 264209, China}

\author{Yang Su}
\affiliation{Purple Mountain Observatory, Chinese Academy of Sciences, Nanjing 210008, China}

\author{Hui Tian}
\affiliation{School of Earth and Space Sciences, Peking University, Beijing 100871, People's Republic of China}

\author{Gang Li}
\affiliation{Department of Space Science and CSPAR, University of Alabama in Huntsville, Huntsville, AL 35899, USA}

\author{Guohui Du}
\affiliation{Shandong Provincial Key Laboratory of Optical Astronomy and Solar-Terrestrial Environment, and Institute of Space Sciences, Shandong University, Weihai, Shandong 264209, China}

\author{Hongqiang Song}
\affiliation{Shandong Provincial Key Laboratory of Optical Astronomy and Solar-Terrestrial Environment, and Institute of Space Sciences, Shandong University, Weihai, Shandong 264209, China}


\begin{abstract}
A complete understanding of the onset and subsequent evolution of confined flares has not been achieved. Earlier studies mainly analyzed disk events so as to reveal their magnetic topology and cause of confinement. In this study, taking advantage of a tandem of instruments working at different wavelengths of X-rays, EUVs, and microwaves, we present dynamic details of a confined flare observed on the northwestern limb of the solar disk on July 24th, 2016. The entire dynamic evolutionary process starting from its onset is consistent with a loop-loop interaction scenario. The X-ray profiles manifest an intriguing double-peak feature. From spectral fitting, it is found that the first peak is non-thermally dominated while the second peak is mostly multi-thermal with a hot~($\sim$10~MK) and a super-hot~($\sim$30~MK) component. This double-peak feature is unique in that the two peaks are clearly separated by 4 minutes, and the second peak reaches up to 25-50~keV; in addition, at energy bands above 3~keV the X-ray fluxes decline significantly between the two peaks. This, together with other available imaging and spectral data, manifest a two-stage energy release process. A comprehensive analysis is carried out to investigate the nature of this two-stage process. We conclude that the second stage with the hot and super-hot sources mainly involves direct heating through loop-loop reconnection at a relatively high altitude in the corona. The uniqueness of the event characteristics and complete data set make the study a nice addition to present literature on solar flares.
\end{abstract}


\section{introduction} \label{sec:intro}

Confined flares, a major group of solar activities, refer to solar flares without association of coronal mass ejections (CMEs). There exists a long-standing interest in understanding the onset mechanism and subsequent energy release process of this group of solar flares.

It has been suggested that there are several factors that determine a solar flare to be eruptive (i.e. accompanied by a CME) or confined (or non-eruptive). According to present scenarios of solar flares, the post-reconnection configuration is important to determine whether a magnetic structure is eruptive or not. In the loop-loop interaction scenario \citep[e.g.,][]{Pall77,Sui06,Kush14,Benz17}, the post-reconnection structure may be simple loops, then the flare is unlikely to erupt due to a lack of free magnetic energy. In other scenarios such as the breakout model \citep[e.g.][]{Antioc1998,Antioc1999} \citep[see,][for latest observational reports]{Chen16,Chen17}, and loss-of-equilibrium flux rope model \citep[e.g.,][]{Forbes1911,LinJ00,Low01,Toro&Klie05,Chen07}, the post-reconnection structure may contain a well-developed twisted flux rope. In this case, the effect of overlying arcade is important to determine whether the eruption is successful or failed \citep{JiHS03}. For instance, if the magnetic field of the overlying arcade is strong enough to stop the tentative eruption of the underlying flux rope, then the eruption becomes failed and thus confined.

Most earlier observational studies on confined flares analyzed disk events so as to reveal the magnetic configuration and thus to investigate the magnetic origin of confinement \citep[e.g.,][]{Yang14,Zucc17}. With a comparison study on the basis of a small sample of X-class confined and eruptive events, \cite{Wang07} proposed that the confined flares tend to occur at the center of the active region while eruptive events tend to occur close to the outer border of the active region. This is consistent with the above analysis on the effect of overlying arcades. \cite{Yang14} also reached a similar conclusion that overlying loops play an important role in a failed eruption. These observational results are consistent with those deduced from numerical simulations \citep[see, e.g.,][]{Toro&Klie05,Klie&Toro06}. From these and subsequent numerical studies, the concept of decay index, representative of the declining gradient of magnetic field strength of the background field has been proposed, and used in observational studies to infer the possibility of flux rope confinement \citep[e.g.,][]{Wang2015NC}.

Despite significant progress made, a complete understanding of confined flares has not been achieved. Observations of disk events are useful in analyzing the magnetic topology of the event, yet with disadvantages in revealing the details of loop-loop interaction during the flare. This is mainly due to the projection effect and the bright background emission from the associated active region. On the other hand, limb events have the advantage that the loop structure may be directly projected onto the dark plane of sky, allowing one to clearly view their dynamic evolution. This study takes the advantage by presenting a well-observed limb event. Many dynamic details of loop-loop interaction during the onset and subsequent evolutionary stage can be revealed.

In the classical picture of solar flares \citep[also called the CSHKP scenario;][]{Carm64,Stur66,Hira74,Kopp76}, energetic particles are accelerated through reconnection taking place in the corona. These energetic particles then flow along the post-reconnection field lines toward the chromosphere. Their interaction with dense chromospheric plasmas results in HXR sources, together with their thermalization. The chromospheric plasmas can be heated up to 10-20~MK. A strong pressure gradient along the post-reconnection field line is then present, which accelerates the heated plasmas upward to fill the post-reconnection field line. This is the well-known chromospheric evaporation process \citep[e.g.][]{Neup68,Tian15,LiY15}. These heated plasmas emit in SXRs and are constituents of post-flare loops.

According to the above picture, the 10-20~MK hot plasmas are not heated directly at the primary energy release region. Depending on the time taken by the chromospheric evaporation process, the peak of the SXR profiles may get delayed relative to that of the HXR profiles. Indeed, earlier studies have found double X-ray peaks of a solar flare \citep{Kane70,LiYP09} with the first peak being impulsive and non-thermally dominated while the second peak is mainly thermal and more gradual. According to \cite{LiYP09}, the delay is less than one to two minutes, and the double-peak feature is mainly revealed in the energy range of 10-25~keV, while not visible in lower or higher energy bands. Here in our event, a double-peak X-ray feature is observed. The feature is similar to earlier reports in the manner that the first peak is non-thermally dominated and the second one is more gradual and thermally dominated, yet with other significant unique characteristics, such as the very long temporal separation of the two peaks, and the very high energy level (25-50~keV) of the second peak. Recognizing and understanding the two HXR peaks of the confined flare are the major motivation of the present study.

\section{Observational Data and Event Overview} \label{sec:Data&Ov}

\begin{figure}[p]
\includegraphics[width=16cm]{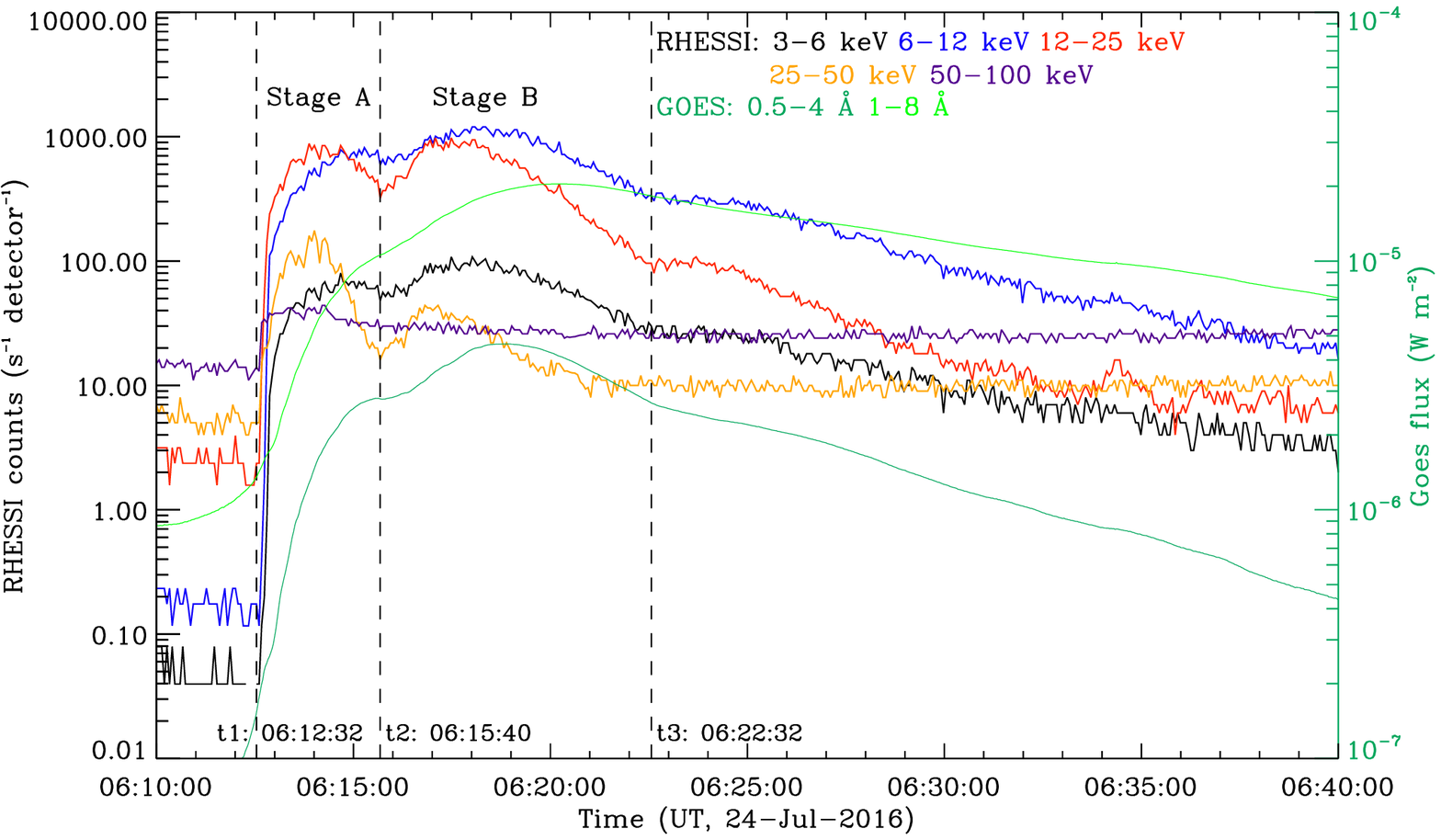}
\caption{GOES and RHESSI X-ray light curves of the M2-confined flare. According to the two peaks, the event is divided into two stages: Stage A and Stage B. The time t1 represents the onset time of the flare (also the starting time of Stage A), t2 the ending time of Stage A (also the starting time of Stage B), and t3 the ending time of Stage B.} \label{Fig1}
\end{figure}

According to the GOES soft X-ray light curve (Figure~\ref{Fig1}), the class of the flare is M2.0. The flare is observed from the northwestern limb of the solar disk. It is originated from the NOAA Active Region (AR) 12567, starting at 06:12~UT and peaking at 06:20~UT on July 24th, 2016. The flare is well-observed by a tandem of space- and ground- based instruments across a wide range of wavelengths, including the Atmosphere Imaging Assembly \citep[AIA;][]{Leme12} onboard the Solar Dynamics Observatory \citep[SDO;][]{Pesn12} at various UVs and EUVs, the Ramaty High Energy Solar Spectroscopic Imager \citep[RESSI;][]{Hurf02,LinR02} at SXRs and HXRs, and the ground-based Nobeyama Radioheliograph \citep[NoRH;][]{Naka94,Taka97} at microwaves.

RHESSI observes solar X-ray and gamma-ray emission above 3~keV with high cadence ($\sim$4~s), spatial resolution ($\sim$3~arcsecs), and energy resolution ($\sim$1~keV). The AIA has the capability of imaging plasma structures at different temperatures from 20,000~K to over 20~MK, with high spatial (0.6~arcsecs pixel size) and temporal (12~s) resolutions to image the solar atmosphere in 7 EUV passbands. The NoRH can provide imaging data at both 17 and 34~GHz, with a 1-s cadence (up to 0.1~s in event mode).

As shown in Figure \ref{Fig1}, two distinct peaks are present in the RHESSI X-ray light curves. These two peaks can be observed from 3 to 50~keV, with a 4-minute long time delay between them. The long time separation and the broad energy range are the most notable characteristics to distinguish this event from those reported in earlier studies \citep[e.g.,][]{LinR76,Sui07,LiYP09}.

The first peak starts at 06:12:32~UT (referred to as t1). At this time, X-ray flux of the whole energy band (3-100~keV) rises impulsively, and peaks at around 06:13~UT, while the 3-12~keV flux peaks 1 minute later. These characteristics are consistent with the scenario of chromospheric evaporation \citep{Neup68,LiYP09}. At around 06:15:40~UT (t2), the X-ray fluxes of all energy bins decrease to minimal values. This indicates the termination of the first peak. After t2, the fluxes of 3-50~keV start to increase again, leading to the second peak. Comparing to the first one, the second peak grows more slowly and lasts longer ($\sim$7~minutes). The X-ray fluxes of higher energy range peak around 06:17~UT, and the fluxes of lower energy (3-12~keV) peak 1 minute later. Around 06:22:32~UT (t3), the X-ray fluxes of all energy bins decrease to local minima, indicating the termination of the second peak.

According to the evolution of the X-ray curves described above, in particular, the two-peak feature, this event can be divided into two stages, Stage A and Stage B. In Stage A, the flare is triggered around t1, and the X-ray fluxes start to increase sharply, and then get to the first peak. At t2, the X-ray fluxes decrease to the minimal value and then start to increase immediately again, indicating the end of Stage A and the start of Stage B. The physical process giving rise to the unique second X-ray peak is the most intriguing part of this event.

\begin{figure}
\includegraphics[width=16cm]{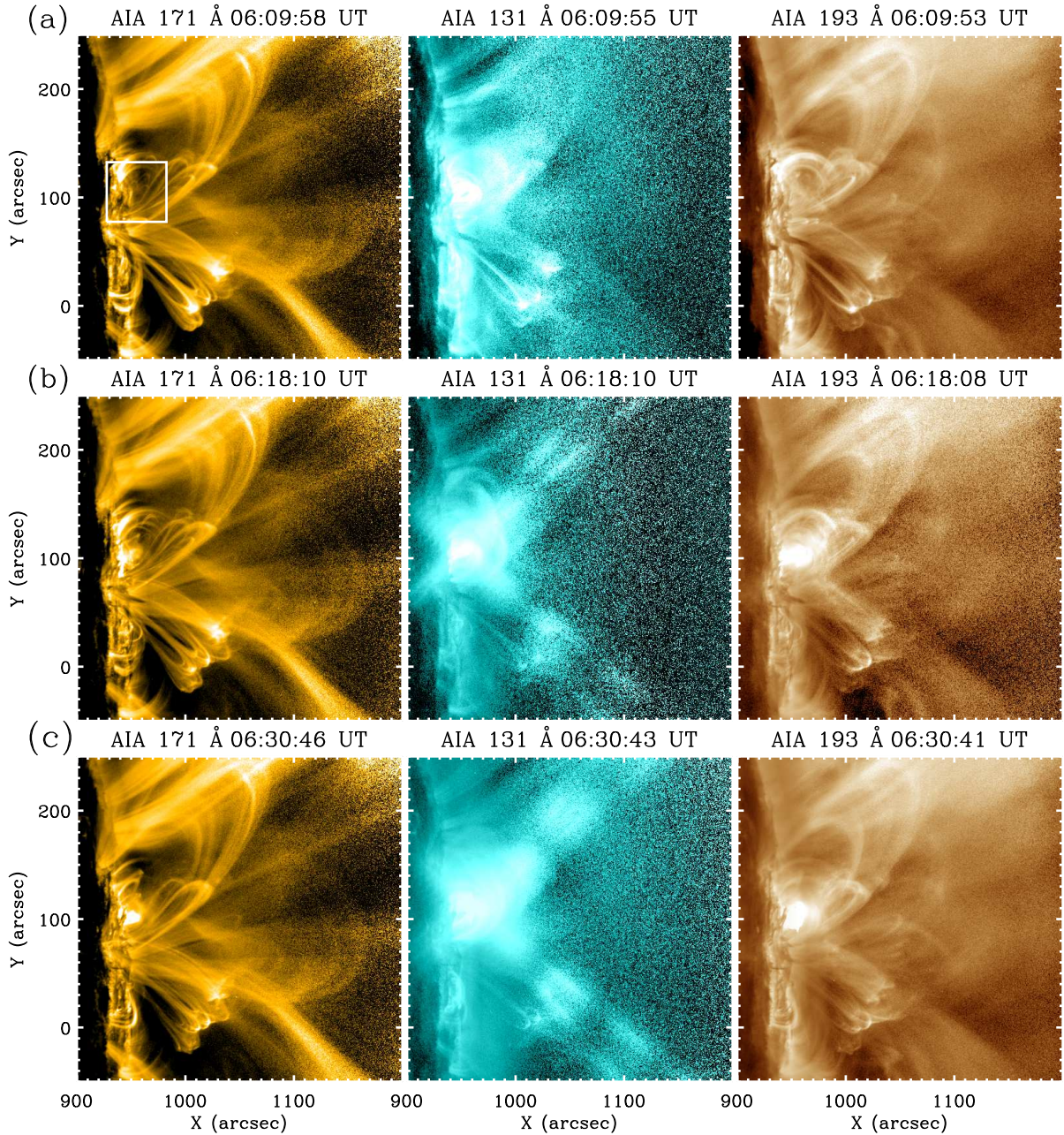}
\caption{AIA images for the whole flare region observed in three channels, 171, 131 and 193~\AA{}. The white box represents the major flare region. Panels (a), (b) and (c) show the images at the start, peak and end of this event, respectively.
\protect\\(The online animation of this figure shows evolution in the three channels shown above plus the 211 \AA{} channel.)}\label{Fig2}
\end{figure}

Figure \ref{Fig2} and the accompanying animation show the evolution of AIA-observed EUV structures in the whole flare region during this event. From the animation, we did not observe any signature of opening of large-scale arcade and ejection of materials, hence the classification of this event is concluded to be a confined flare. As shown in Figure \ref{Fig2} and the animation, several loops become very bright, and then cross each other rapidly. Note that the footpoints of the flare loops are partially occulted by the disk. This allows one to observe the faint X-ray and microwave sources in the corona. A very bright EUV source appears in high-temperature channels (131 and 193~\AA{}) at the peak time of this flare at the center of the core flare region. It has been indicated with the white box plotted in Figure~\ref{Fig2}(a). The size of the source is about $20\times30$ arcsecs$^{2}$. It is more evident in 193 ($\sim$1.6~MK and 18~MK) and 131 \AA{} ($\sim$10~MK) images, but not clear at 171 ($\sim$0.6~MK) and 211 \AA{} ($\sim$2~MK). This indicates that the emitting plasmas are at a very high temperature ($>10$~MK). The dynamical evolution of the loops and the morphological change of the bright EUV source also show the two-stage feature as observed with the X-ray light curves. At the end of the flare, post-flare loops are observed in AIA 171~\AA{} images, as shown in Figure~\ref{Fig2}(c).

\section{Analysis of EUV, X-ray, and Microwave Data} \label{sec:analysis}

We used data observed at various wavelengths, including EUV from AIA/SDO, SXR-HXR from RHESSI, and microwaves from NoRH, to investigate the detailed evolution of the
event. We found that the two-stage characteristic, as defined above with the X-ray light curves, is also present in the EUV and microwave observations. This strongly indicates that the event is characterized by an intrinsic two-stage energy release process.

EUV emissions are from thermal plasmas at different temperatures. With the knowledge of response functions at various passbands, they can be used to infer plasma temperatures. This study mainly focuses on AIA passbands with significant high-temperature responses, such as those at 94, 131 and 193~\AA{}. Note that both the 131 and 193~\AA{} passbands have dual temperature responses, with the high-temperature response peaking at $\sim$10~MK for 131~\AA{} and $\sim$18~MK for 193~\AA{}, as mentioned earlier. EUV emissions are also very useful in inferring the coronal magnetic configuration, due to the well-known property that coronal plasmas are effectively frozen into the magnetic field.

During solar flares, X-rays are mainly from thermal or non-thermal bremsstrahlung emissions, while microwaves are mainly from gyro-synchrotron emissions. Thus, they contain complementary information on the underlying energetic electrons and thermal plasmas, such as the energy spectra of non-thermal electrons, temperature and emission measure of thermal plasmas \citep[e.g.,][]{White11}. To expose these pieces of information, imaging and spectral analyses of X-rays and microwaves are necessary. A combined analysis of all the available data sets will be carried out for a complete understanding of the event.

\subsection{Analysis of the EUV Data of the Event} \label{ssec:euv}

In Section~\ref{sec:Data&Ov}, an overview of the general evolutionary process of the flare has been presented. In particular, from the X-ray light curves, we have separated the process into two stages. In this section, to investigate the detailed loop dynamic evolution and magnetic configuration of the flare, we focus on its core region. To better visualize the loop dynamics, we employed the Multi-scale Gaussian Normalization method \citep[MGN;][]{Morg2014} to further process the EUV data. This method allows us to achieve visual enhancements of loop structures in AIA images. It has been frequently used in imaging analysis of solar observational data \citep[e.g.,][]{Fu14,Luna17}.

\begin{figure}
  \centering
  \includegraphics[width=16cm]{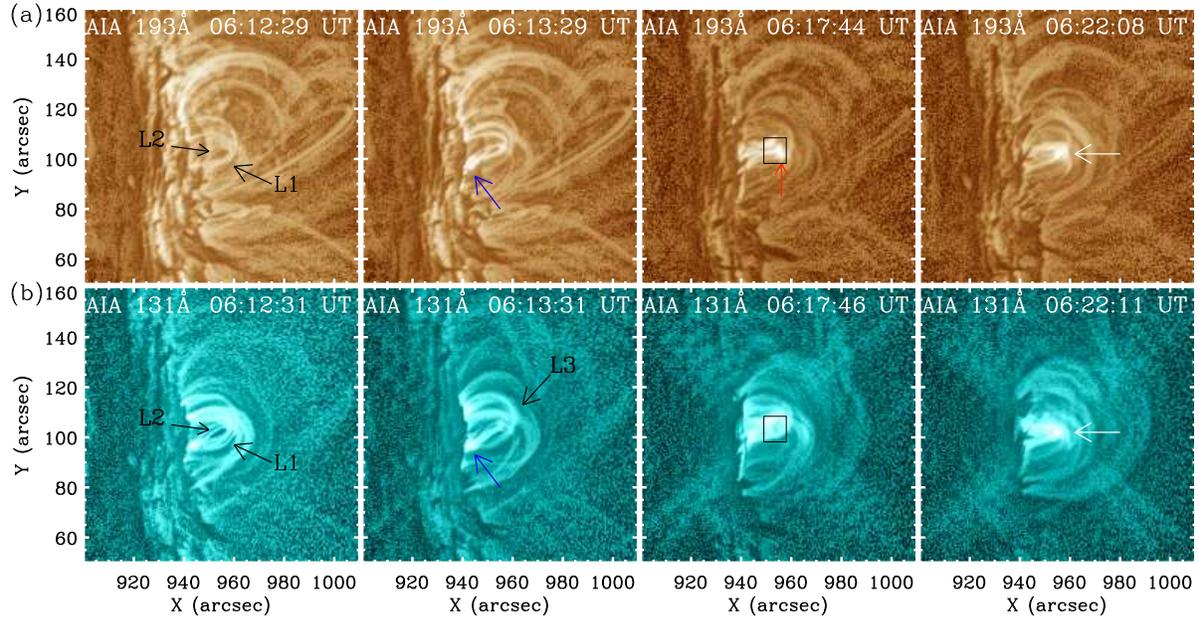}
  \caption{Panels (a)-(b) show the AIA 193 and 131~\AA{} images processed by the MGN method. The four columns show the loop dynamics in the core flare region, representing the onset, Stage A, Stage B, and the post-flare configurations. The black arrows point to the loops L1, L2, and L3. The blue arrows point to the footpoint brightening region, and the red arrow points to the upper part of the loop-loop intersection region, marked by the black box. The white arrows point to the newly-formed loops in the recovery phase. \protect\\(An animation of this figure is available. The evolution of the two channels together with the 94~\AA{} are shown.)}\label{Fig3}
\end{figure}

In Figure~\ref{Fig3} and the accompanying animation, we show the complete evolutionary sequence of the flare from 06:00 UT to 06:30 UT. In general, we can separate the whole process into four distinct phases, including the onset phase, the subsequent two phases corresponding to stages A and B (as defined above with the X-ray light curves), and the recovery phase. In Figure~\ref{Fig3}, we show AIA images that are representative of the dynamical evolution of each phase. In the following paragraphs, we will describe them one by one.

\emph{Onset of the flare.} The onset of the flare is defined to be the short interval before the sharp rise of the RHESSI HXR light curves, from 06:11 to 06:13~UT. As observed from the EUV data in this phase, the following important features should be highlighted. At 94 and 131~\AA{}, a system of loops exists even before the onset of the flare, while at 193~\AA{} the loop system is not observable before 06:11~UT. Later, at 193~\AA{} two sets of loops (L1 and L2, pointed by the black arrows) start to appear and become brighter in the field of view (FOV), gradually. In the meantime, the loop structure at 94 and 131~\AA{} also becomes brighter. In other cooler passbands of AIA, the loop system is not clearly observed. This indicates that the loops contain high-temperature plasmas.

\begin{figure}
  \centering
  \includegraphics[width=16cm]{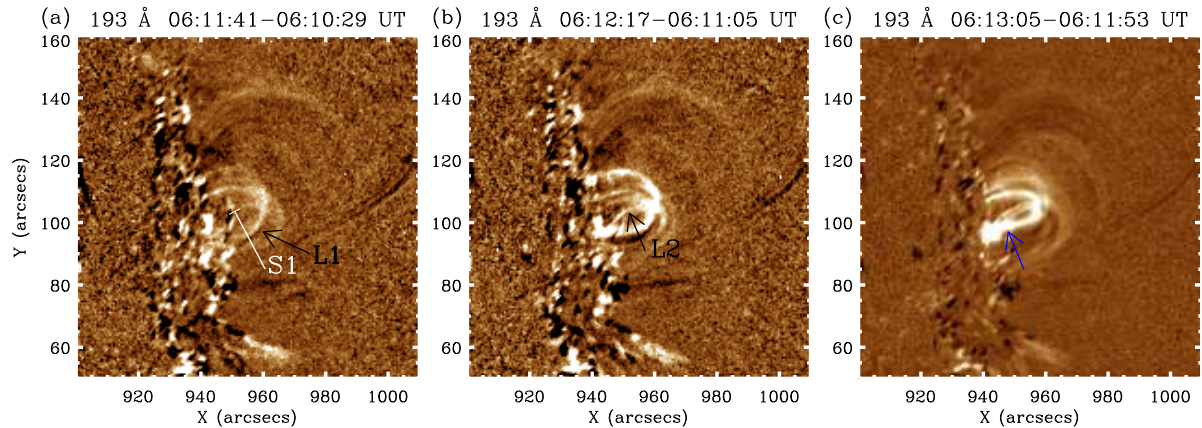}
  \caption{Panels (a)-(c) show the AIA running difference images of the onset stage of the event at 193~\AA{}. The arrows point to the loops L1 (a), L2 (b) and the footpoint brightening (c). The solid line S1 in (a) is the slit for distance-time maps, with the starting point marked by a short dash.  \protect\\(An animation of this figure is available.)} \label{Fig4}
\end{figure}

\begin{figure}
\includegraphics[width=16cm]{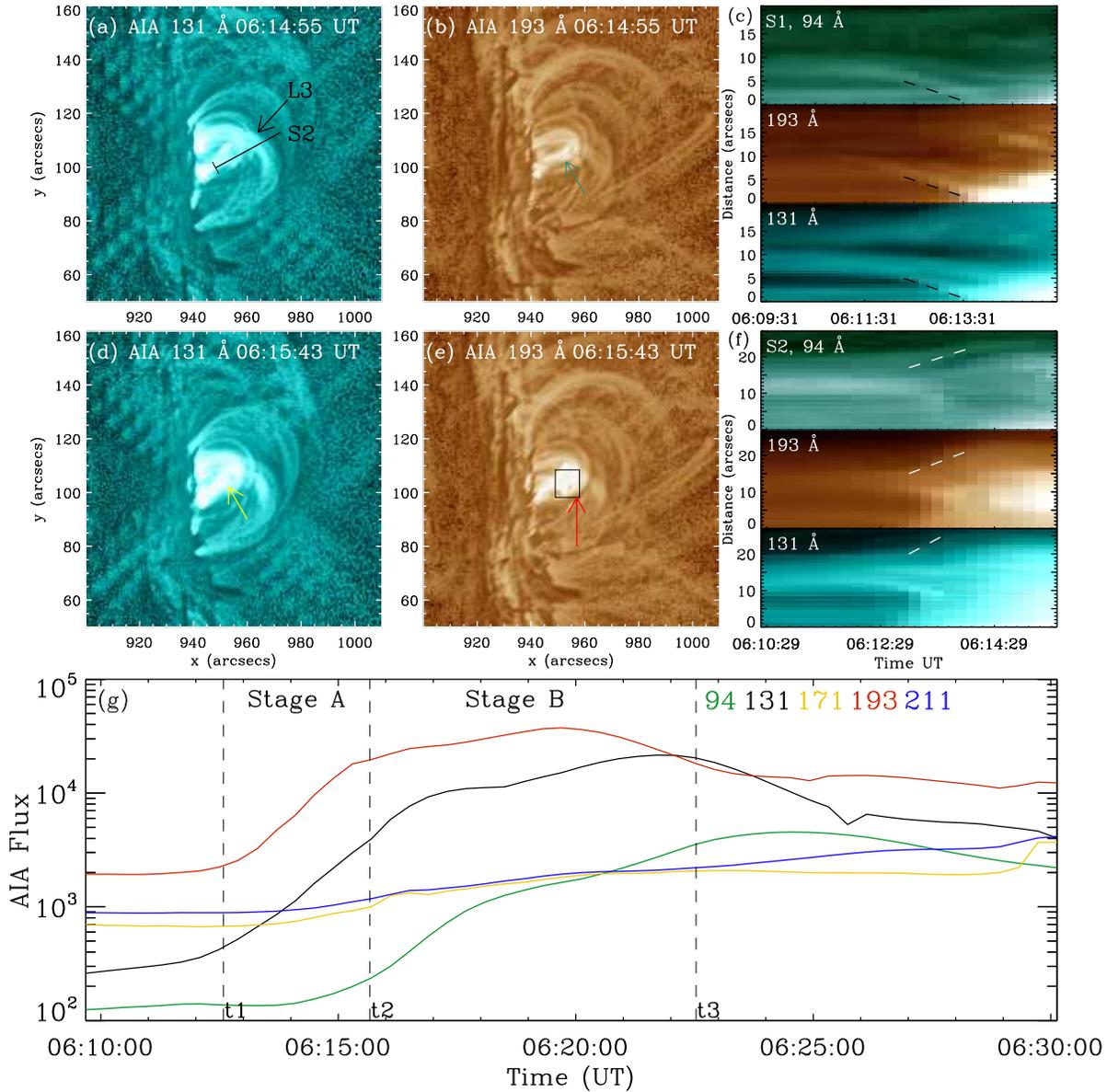}
\caption{Panels (a)-(b) and (d)-(e) show the MGN-processed images after the footpoint brightening in Stage A. The arrows point to the loops L3 (black arrow in (a)), the loop-loop intersection region (green arrow in (b)), the column-like structure (yellow arrow in (d)), and the upper part brightening of the loop-loop intersection region (red arrow in (e)).The solid line S2 in (a) is the slit for distance-time maps, with the starting point marked by a short dash. Panels (c) and (f) present the slice-time plot showing the motion of L1 along S1 (see Figure~\ref{Fig4}(a)), and the motion of L3 along S2, respectively, in 94, 193, and 131~\AA{} channels. The dashed lines are used to estimate the speed of the loop motions. Panel (g) shows the AIA light curves of this event in five channels (94, 131, 171, 193, and 211~\AA{}), averaged over the black box shown in panel (d) (see also Figure~\ref{Fig3}). Fluxes obtained in different channels are plotted in different colors.} \label{Fig5}
\end{figure}

Among all the three passbands, the two sets of loops are clearly observed with a strong dynamic interaction. Especially, they appear to cross each other at an altitude of $\sim20\arcsec$. In addition, the L1 loop manifests a rapid northward motion that is basically parallel to the solar disk. The above dynamic evolution of loops is also clearly observed from running difference images in the above passbands. In Figure~\ref{Fig4}, we present such images and the accompanying animation recorded at 193~\AA{}. In Figure~\ref{Fig5}, we also present the distance-time images along the slice S1 (at 94, 131, and 193~\AA). The speed of the rapid L1 motion can be estimated by linear fit of the distance measurements, which is found to be $\sim$40~km s$^{-1}$. Such speed of loop motion, being parallel to the solar disk at such low altitude (10-20\arcsec{} above the disk), is very fast and rarely reported. Since the loop footpoint could not be observed in this event, the origin of this fast motion remains unclear. Later in Section~\ref{sec:scenario}, a scenario involving footpoint-interchange reconnection will be proposed to understand its origin.

The fast motion of L1 makes the two sets of loops (L1 and L2) tangled together more tightly. A brightening region appears at the lower part of the loop-loop intersection region (pointed by blue arrows). This brightening signifies the start of the flare (and Stage A).

\emph{Stage A.} According to the RHESSI X-ray light curves (25-50 keV), this stage, representative of the impulsive stage of the flare, starts from 06:12:32~UT and ends at 06:15:40~UT. As seen from Figure~\ref{Fig3} with its accompanying animation, and Figure~\ref{Fig5}, this stage is characterised by strong footpoint brightening. Note that the footpoints are partially occulted so the flare level according to the GOES X-ray curves may be underestimated. There exist two major locations of footpoint brightening, which are co-incident with the continuous rapid loop motion and the consequent tightening trend of the loop-loop tangle. In addition to the footpoint brightening, the whole interaction region (pointed by the green arrow) and the associated loops also get brightened significantly at 131 and 193~\AA{}. This indicates a strong heating process.

During this stage, another set of loops (see the black arrow in Figure~\ref{Fig3} and \ref{Fig5}(a)) appears and expands above the L1-L2 intersection region. The upward moving velocity of these loops is estimated to be 40-80~km s$^{-1}$ according to the distance-time analysis (see Figure~\ref{Fig5}(f)). At the end of this stage, the loop-loop intersection part becomes highly tangled with an inclining trend toward the moving direction of L1. This strongly indicates the important role of the rapid L1 motion, not only in the onset but also in this impulsive stage of the flare.

At the end of this stage, a local brightening region appears at 193~\AA{} around the upper part of the loop-loop intersection region. The start of this upper brightening is co-temporal with the second rise of the RHESSI X-ray light curves and signifies the start of Stage B. Note that the brightening structures are clearly observed at both 131 and 193~\AA{}, while they remain not very clear at 94~\AA. This indicates that the temperature of emitting plasmas is higher than the effective response range of the 94~\AA{} passband.

\emph{Stage B.} According to the RHESSI X-ray light curves, this second stage with the second HXR peaks is from 06:15:40 to 06:22:32~UT. As mentioned, at the end of the earlier stage, there appears to be a strong brightening at the upper part of the loop-loop intersection region (see the red arrow in Figure~\ref{Fig5}(e), around 15\arcsec{} above the solar disk). The brightening reaches the maximum around 06:16:11~UT, and causes the saturation of the AIA-193 passband. At 131~\AA{}, the brightening region is a longer and highly-inclined column-like structure, which extends over the whole intersection region (see the yellow arrow).

The upper part of the loop-loop intersection region declines rapidly in brightness at both 131 and 193~\AA{} after 06:16~UT. It starts to become brighter again from 06:18~UT, and reaches the maximal brightness around 06:19~UT. This causes the saturation at 193~\AA{}, again. Later, its brightness declines rapidly. The two episodes of brightening around the upper part of the loop-loop intersection region represent the major characteristic during this stage.

The bright EUV source is located high in the corona, while the region below the source is much weaker in brightness. Around 06:22~UT, a set of newly-formed bright loops appear at 131 and 193~\AA{} (see the white arrows in Figure~\ref{Fig3}). This indicates the end of Stage B. It should be noted that at the time of the second X-ray peak ($\sim$06:18~UT), the EUV source is the brightest at 193~\AA{} among other available AIA passbands including the 131 and 94~\AA{}. This indicates that the source contains plasmas with temperature at the level of the effective response range in the 193~\AA{} passband (peaking at 18~MK).

In Figure~\ref{Fig5}(g), we plot the EUV fluxes (in arbitrary unit) of the five passbands, averaged over the upper part of the L1-L2 intersection region (see the black box). All flux profiles manifest the two-stage feature, consistent with what has been defined using the RHESSI X-ray light curves. Stage A is characterized by rapid increases in fluxes for high-temperature channels (94, 131, 193~\AA{}), while in cooler passbands (171 and 211~\AA{}) the fluxes do not increase considerably.

Stage B is characterized by gradual and continuous increase in fluxes in the above high-temperature passbands, and the fluxes in cooler passbands are observed with slight increase. The 193~\AA{} flux curve reaches its peak at 06:20~UT, and the peak of the 131~\AA{} curve is reached about 2.5 minutes later. This continuous increase in fluxes indicates a heating process, while the temporal delay of flux peaks indicates a subsequent cooling process.

In summary, the EUV emissions in the two stages are distinct from each other. During Stage A, the EUV sources are mainly located at the footpoint-loop regions, while they move to a higher altitude without significant footpoint counterpart during Stage B. From the above analysis, we suggest that Stage A represents a typical impulsive phase of a solar flare, while Stage B represent a subsequent heating process occurring high in the corona.

\subsection{Imaging and Spectral Analysis of RHESSI X-Ray Data}\label{ssec:xray}

\begin{figure}
\includegraphics[width=16cm]{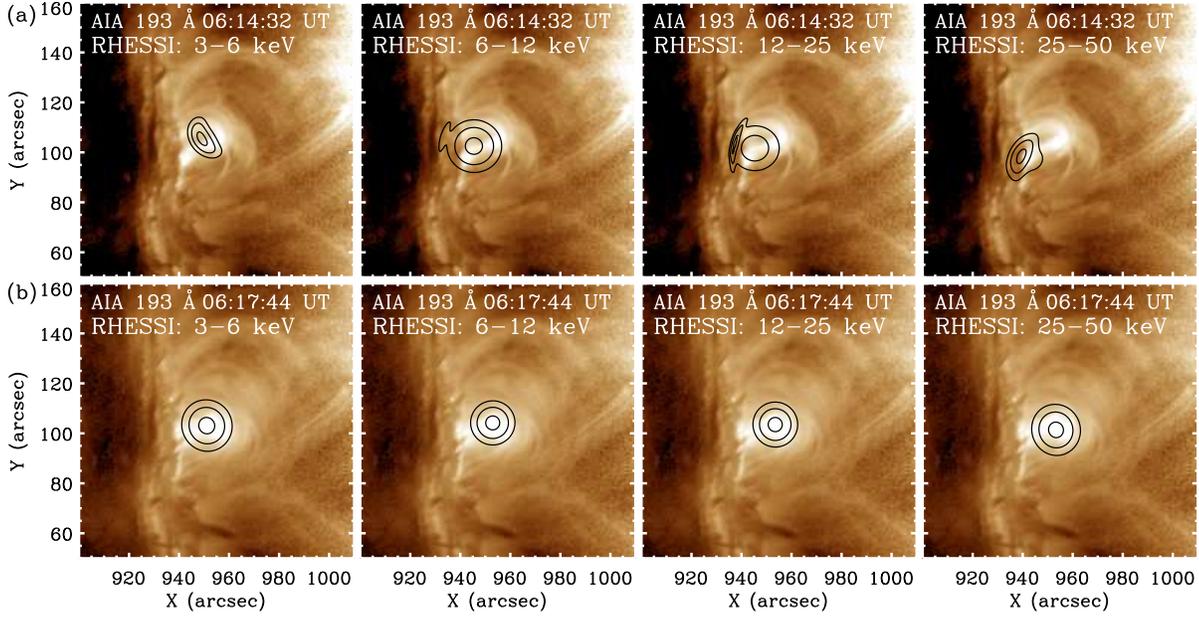}
\caption{AIA images (193~\AA{}) overlaid with RHESSI X-ray contours observed during the two stages (upper panels for Stage A, and lower panels for Stage B). Contour levels are given by 90\%, 95\% and 99\% of the maximal flux in each energy band (3-6, 6-12, 12-25, 25-50 keV), reconstructed using the PIXON algorithm with data from detector 3 and 8. The time intervals of reconstruction are 06:12:02-06:16:00~UT and 06:16:00-06:20:00~UT.}\label{Fig6}
\end{figure}

Figure~\ref{Fig6} shows the X-ray sources superposed onto the AIA images. It can be seen that the energy dependence of source centroid during Stage A and B is very different from each other.

During Stage A, the source centroid distances present a clear energy dependence, with those at higher energy bands being located lower. For instance, the 25-50~keV source centroid is very close to the solar limb, co-spacial with the bright EUV footpoints, while the centroid of the 3-6~keV source is 10\arcsec{} above the disk, near the center of the loop-loop intersection region.

During Stage B, the centroid distances of X-ray sources do not present an observable energy dependence. For different energy bands, the sources are all located around 15\arcsec{} above the disk. In other words, they are emitted from the same location, co-spacial with the very-bright high-temperature EUV sources observed at 193 and 131~\AA{}. This indicates that the X-ray and bright EUV sources at this stage have the same origin.

\begin{figure}
  \centering
  \includegraphics[width=16cm]{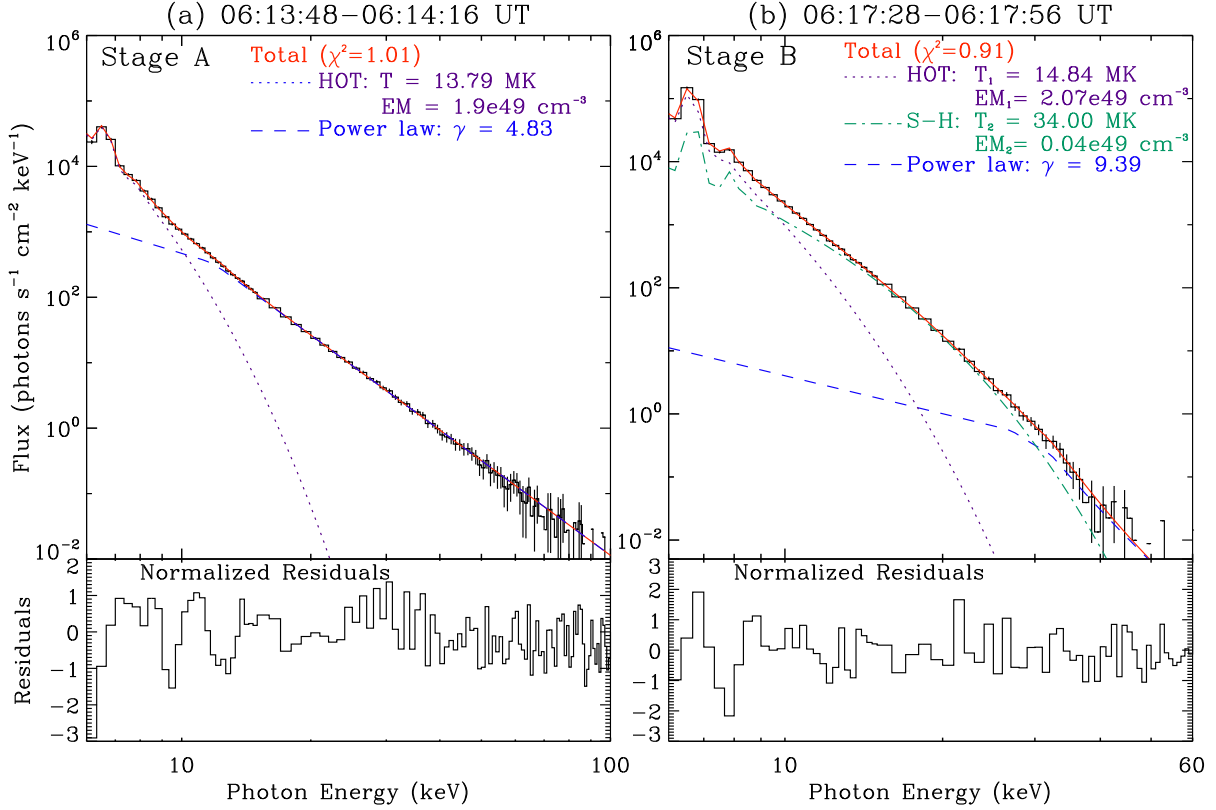}
  \caption{RHESSI spectral fitting results of the two HXR peaks. Detectors 3 and 8 are used. Panel (a) shows the spectral fitting result of the first peak, and panel (b) shows the spectral fitting result of the second one. The histograms with error bars represent the background-subtracted spectra. Thermal components (purple and green lines) and non-thermal components (blue lines) are plotted. The total fit is given in red. The normalized residuals are shown at the bottom.}\label{Fig7}
\end{figure}

To infer the properties of the thermal plasmas and non-thermal electrons accounting for X-ray emissions, we carried out spectral fittings of RHESSI data using the standard OSPEX software distributed with the SolarSoftWare package. The results for the intervals around the two X-ray peaks are shown in Figure~\ref{Fig7}.

The spectrum for the first peak in Stage A can be well-fitted with two components of electron distribution, including a thermal component (purple dotted line) and a power-law component (blue dashed line). The Chi-square of the fit is 1.01. The temperature of the thermal component is 13.79~MK, the emission measure is $1.9\times10^{49}$~cm$^{-3}$, and the spectral index of the power-law component is -4.83. Above 10-15~keV, the spectrum is dominated by the non-thermal component. This represents a typical spectrum observed during the impulsive stage of solar flares.

The major part of the spectrum for the second peak in Stage B (below 40-50~keV) is much softer in comparison to that of the first peak. This part can be well-fitted with two thermal components, including one hot (H) component and one super-hot (SH) component (Figure~\ref{Fig7}(b)). The H component has a temperature of 14.84~MK and an emission measure of $2.07\times10^{49}$~cm$^{-3}$, and the SH component has a temperature of 34 MK and an emission measure of $0.04\times10^{49}$~cm$^{-3}$. A third power-law component, very soft with a power law index at -9.39, is required to fit the high energy end of the spectra ($>$40~keV). The uncertainty is large at high energy due to lower photon count.
The Chi-square of this fitting is 0.91.

To infer the temporal evolution of the X-ray spectra during Stage B, we further divide Stage B to 5 intervals and perform the spectral analysis for these five intervals separately. We employed the same H-SH thermal components plus a very-soft high-energy power law spectra for the fittings. The results are shown in Table~\ref{Tab:1}. The temperature of the H component is around 13-15~MK, and that of the SH component varies from 25-35~MK. Comparing these parameters at different intervals, we infer that the plasmas experience a first-heating-then-cooling process. This is consistent with the result deduced using the flux curves recorded with AIA 193-131~\AA{} passbands. In addition, the spectral indices of the power-law component, as well as the corresponding break energy, increase with time. This indicates that the non-thermal component becomes more and more negligible. At the end of this stage (the last interval in Table~\ref{Tab:1}), the non-thermal component is not necessary anymore.

\subsection{Imaging and Spectral Analysis of NoRH Microwave Data}\label{ssec:mw}

\begin{figure}
  \centering
  \includegraphics[width=16cm]{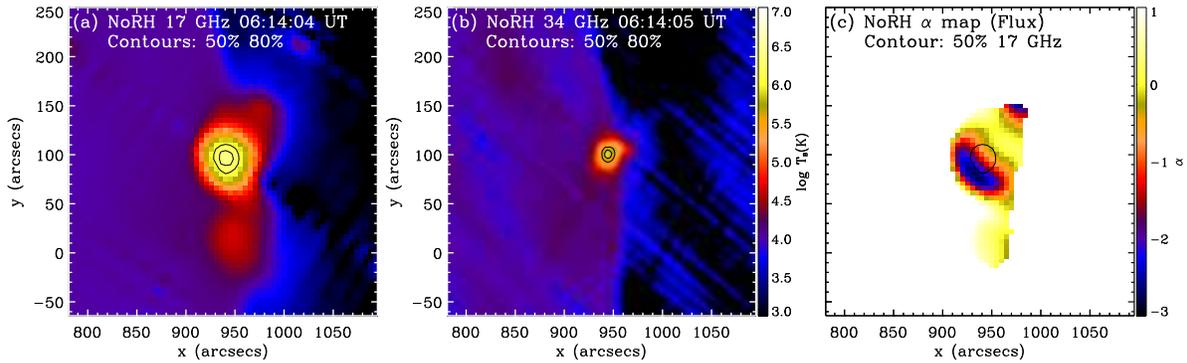}
  \caption{Panels (a)-(c): NoRH microwave images of $T_B$ at 17, 34~GHz, and the deduced map of flux-density spectra ($\alpha$) at the flare peak time, overlaid with the 50 and 80\% contours of the corresponding maximum $T_B$. \protect\\(An animation of this figure is available.)}\label{Fig8}
\end{figure}

\begin{figure}
  \centering
  \includegraphics[width=16cm]{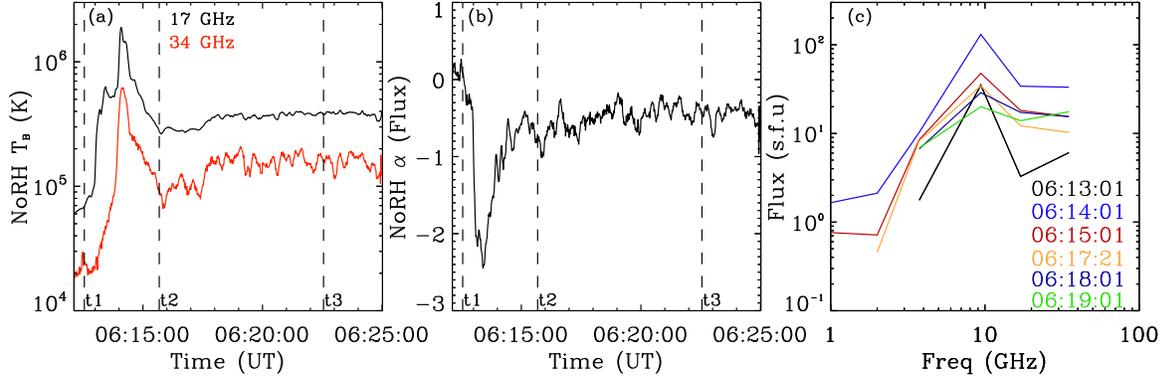}
  \caption{Temporal profiles of the mean $T_B$ (a) and spectral indices (b) averaged within the corresponding 50\% contour as shown in Figure~\ref{Fig8}(a)-(c). Panel (c) shows NoRP flux densities from 1 to 35~GHz at several times.}\label{Fig9}
\end{figure}

\begin{figure}
  \centering
  \includegraphics[width=16cm]{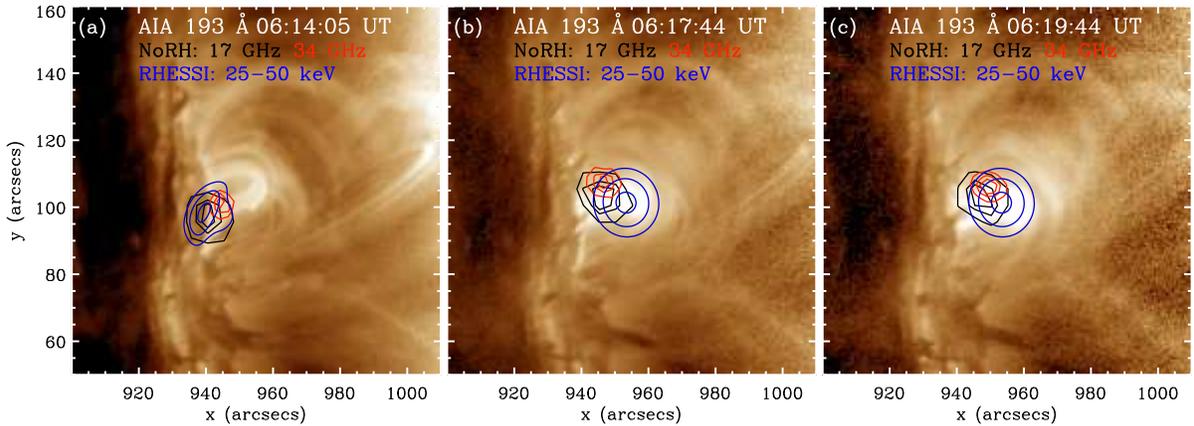}
  \caption{AIA 193~\AA{} images ((a) for Stage A, and (b)-(c) for Stage B), superposed with the 50 and 80\% contours of the maximal $T_B$ at relevant frequency, and the 90, 95, and 99\% contours of the RHESSI X-ray data of 25-50~keV.\protect\\(An animation of this figure is available.) }\label{Fig10}
\end{figure}

The microwave images at 17 and 34~GHz are recorded by NoRH at a temporal resolution (1s) much higher than that of the RHESSI X-ray data. The images are shown in Figure~\ref{Fig8} and the accompanying animation. In Figure~\ref{Fig9} we plot the temporal profiles of $T_B$ averaged within the area given by the corresponding 50{\%} contour, the microwave spectral indices given by the 17 and 34~GHz data, and the spatially unresolved spectra given by Nobeyama Radio Polarimeters (NoRP) \citep{Tori79}. Figure~\ref{Fig10} shows the 50 and 80\% contours of the maximum brightness temperature ($T_B$) at relevant frequency, and the 90, 95, and 99{\%} contours of the RHESSI X-ray data of 25-50~keV, superposed onto the AIA 193~\AA{} images. An accompanying animation is also available.

As observed from these figures and accompanying animations, the microwave data also present an evident ``two-stage'' evolution. During Stage A, the temporal variations of $T_B$ at both 17 and 34~GHz are very similar to that of the HXR data. After the onset of the flare, the values of $T_B$s at 17 and 34~GHz increase rapidly from 0.06~MK and 0.02~MK at 06:12~UT to their maxima of 2~MK and 0.6~MK at 06:14~UT, and then decline abruptly to local minima of 0.3~MK and 0.07~MK at 06:15:40~UT, respectively. This is consistent with the termination of Stage A at this time. After 06:16 UT and during the whole Stage B, the $T_B$s only present a slight and gradual increase (in an oscillating manner), to $\sim$0.4~MK at 17~GHz and $\sim$0.2~MK at 34~GHz.

During Stage A, the sources at the two frequencies are very close to the hard X-ray sources, and are co-spatial with the bright loop footpoint (see Figure~\ref{Fig10}(a)). During Stage B, there appears to be a slight rise of the microwave sources. At the time of the second HXR peak (06:18~UT), the microwave and HXR sources are also very close to each (see Figure~\ref{Fig10}(b)). Later (06:20~UT, still during Stage B), different sources basically overlap with each other and with the AIA-observed bright 193~\AA{} source (see Figure~\ref{Fig10}(c)).

We also deduced the temporal evolution of the microwave spectra ($\alpha$) of flux density using the NoRH data, to further investigate the nature of the emission. The result is shown in Figure~\ref{Fig9}(b). We see that during Stage A, $\alpha$ first decreases rapidly from 0.2 to -2.5 at 06:13:30~UT, and then increases back to $\sim$-0.5 (at 06:15~UT). Later, during Stage B, $\alpha$ does not change significantly and mainly remains close to -0.5.

The spatially unresolved data from NoRP at several other frequencies (smaller than 10~GHz) can be used to infer the peak frequency of the total microwave spectrum. The data are shown in Figure~\ref{Fig9}(c). We see that the turnover frequency during the interval of interest is always below or around 10~GHz. This means that the 17-34~GHz data shown here are optically thin throughout the event.

From the above analysis, in particular, the values and temporal evolution of $T_B$ and $\alpha$, we conclude that the microwave emission during Stage A is mainly caused by the optically-thin non-thermal gyro-synchrotron emission \citep[see, e.g.,][]{Dulk85,WuZ16}, while during Stage B the microwave emission is mainly caused by thermal bremsstrahlung. This is consistent with the analysis on RHESSI X-ray data.

\section{A Possible Scenario for the Onset and Two-Stage Process of the Confined flare}\label{sec:scenario}

As mentioned, the whole process of the confined flare can be separated into four phases, including the onset that is characterized by rapid loop motion and loop-loop intersection, Stage A and Stage B with the double X-ray peaks, microwave sources, and significant loop brightening, and the recovery phase.

\begin{figure}
  \centering
  \includegraphics[width=16cm]{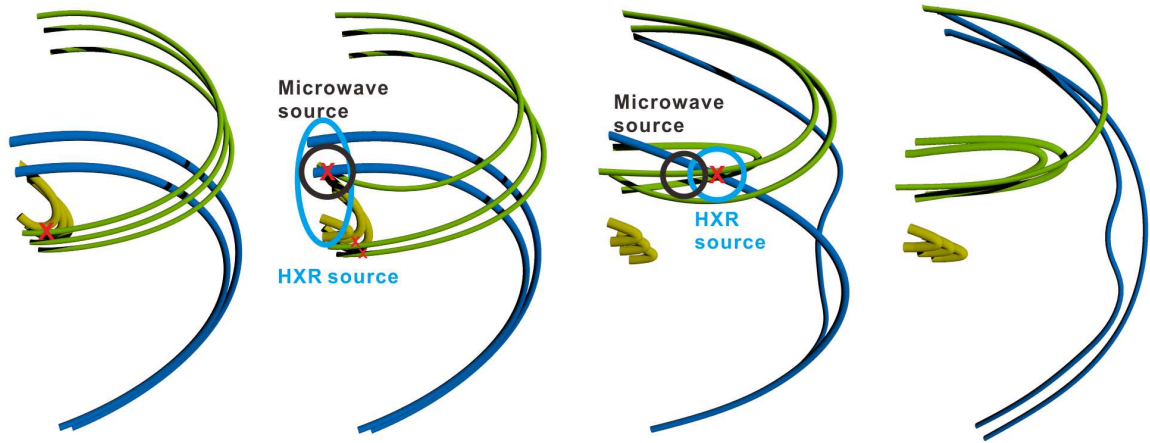}
  \caption{Four schematics of magnetic skeleton, representing the configuration of the four flare phases (in accordance with Figure~\ref{Fig3}). Green and blue loops represent L1 and L2, and red crosses represent the reconnection points. Cyan and black ellipses represent the X-ray and microwave sources. }\label{Fig11}
\end{figure}

In Figure~\ref{Fig11}, we present schematics of magnetic skeleton to explain the flare evolutionary process. Green and blue loops represent L1 and L2, and red crosses show sites of reconnection. From the AIA data, we observed the rapid motion of L1 at a speed of $\sim$40~km~s$^{-1}$. This speed (almost parallel to the solar disk) is too fast to be driven by the slow footpoint motion on the photosphere. To explain its origin, we suggest that there exists a set of low-lying loops (marked in yellow), either encounter being newly emergent or pre-existing, may undergo outward expansion and meet the green loops at a low altitude. This may have triggered reconnection between them. The reconnection results in footpoint exchanges of L1 and the yellow loops, and consequently the rapid motion of L1. It should be noted that the yellow loops may have been occulted by the solar disk and not observable by AIA.

As observed with AIA, the rapid motion of L1 plays an important role in subsequent energy release of the flare. With the rapid motion, the L1 and L2 loops are tangled more and more tightly, first at the lower part and then along the whole region of intersection. This may have triggered reconnection between L1 and L2, first at the lower part, then along the whole region of intersection, and later converging at the upper part of the intersection. As evidenced from the data analyzed above, this L1-L2 reconnection evolves in accordance to a two-stage process. This means that the reconnection quenches for a short interval between Stage A and Stage B. The process has been illustrated in the middle two schematics, corresponding to the double HXR peaks, i.e., the two stages A and B. The locations of HXR and microwave sources during the two stages have been marked with the cyan and black ellipses, respectively. The last schematic presents the recovery phase, showing post-reconnection low-lying and overlying loops.

\section{Conclusions and Discussion} \label{sec:discussion}

In this paper, we presented a multi-wavelength data analysis of a confined M2-class flare, observed on the northwestern limb. The data include those from various EUV channels of AIA/SDO, SXRs and HXRs from RHESSI, and microwaves from NoRH. The flare is of particular interest because of its intriguing double peaks observed in HXRs. The double HXR-peak feature is unique in that there exists a 4-minute delay that is relatively long in comparison with earlier reports of similar events. In addition, the second peak reaches up to 25-50~keV, and between the two peaks the X-ray fluxes in almost all RHESSI energy bands ($>$3~keV) decline significantly.

According to the RHESSI-observed X-ray light curves, we separate the major energy release process into two stages (Stages A and B). The multi-wavelength data analysis shows that both EUV and microwave data manifest this two-stage characteristic. We found that the dynamic evolutionary process of this flare starting from its onset can be well-understood with a loop-loop interaction scenario. From EUV data, the event starts from a rapid motion of coronal loop, which intersects with another set of loops. The rapid loop motion and consequent tighter loop-loop intersection play important roles in the flare onset and energy releases of subsequent stages. Significant brightening takes place first at the lower part, and then extends along the whole intersection region, and later converges on the upper part. This evolutionary sequence is consistent with what we observed in X-rays and microwaves, indicating that reconnection and significant energy releases take place around relevant locations.

It is found that the first HXR peak (Stage A) is non-thermally dominated, typical for the impulsive stage of a solar flare, while the second HXR peak (Stage B) is mostly multi-thermal including one hot~($\sim$10~MK) component and one super-hot~($\sim$30~MK) component. In earlier reports, similar double HXR peaks (the first one non-thermally dominated being harder and more impulsive and the second one thermally-dominated being softer and more gradual) have been interpreted as the consequence of a delayed effect of chromospheric evaporation \citep[e.g.,][]{LiYP09}. Here, in our event, as discussed above, the temporal delay between the two HXR peaks is as long as 4 minutes, longer than that of most events reported earlier ($\sim$1-2~minutes). In addition, the X-ray energy of the second peak can reach up to 50~keV, and the spectral fitting yields that the temperature of the super-hot component is larger or around 30~MK. It is unlikely that a chromospheric evaporation process can heat plasmas to temperatures high enough to emit in this energy band \citep[e.g.][]{Casp10,Casp14}. Furthermore, X-ray and microwave sources, together with the major brightening region at 193\AA{} observed by AIA, all are located at the upper part of the loop-loop intersection region, without counterpart around the footpoint. We thus suggest that the enhanced X-ray and high-temperature EUV emissions are from direct heating of reconnection there, rather than a result of the chromospheric evaporation process.

From this study, it can be inferred that the reconnection in the corona can yield very different physical outcomes. During the first stage (A) of this event, the reconnection mainly occurs around the lower part of the loop-loop intersection, generating non-thermal energetic particles which release HXRs with energy up to 100~keV. On the other hand, during the second stage (B), the reconnection mainly convert magnetic energy into plasma thermal energy. This can be seen from the very-flat microwave spectra, the strong brightening at 193~\AA{}, and the HXR spectra. At this stage, the reconnection takes place at a higher altitude, with magnetic configuration different from that of the first stage, and the plasmas involved are already flare-heated to a temperature of 10~MK, therefore the plasma $\beta$ may be much larger than the value during the reconnection of Stage A. These different plasma and magnetic conditions may account for the different outcome of reconnection.

The two-stage evolutionary process of a confined flare reported here is different from those events with the so-called late phase phenomena \citep{Wood11}. Those events are mainly associated with secondary heating process. For example, \cite{LiuK15} studied several flares with failed eruptions and obvious EUV late phase. In those events, the second peak in the flare emission is mainly observed in cooler passbands (such as 335~\AA{}) of SDO/EVE \citep{Wood12}. They argued that the late phase results from secondary heating induced by reconnection between the failed-to-erupt flux rope and nearby large scale magnetic arcade. In our event, the two-stage energy release process is consistent with the traditional scenario of loop-loop interaction for confined flares, without the presence of a flux rope structure.

Super-hot component of coronal plasmas has been reported in solar flares, first by \cite{LinRP81}. Latest studies suggest that this component is different in physical origin from the usual hot component (10-20~MK), with the super-hot component being generated through direct heating via reconnection that occurs high in the corona, and the hot component through the chromospheric evaporation process \citep[see, e.g.,][]{Casp14,Kruc14}. Our study on the origin of the super-hot component is consistent with these earlier studies in the regard that it is associated with direct heating of reconnection. Yet, this component appears only during the second stage of a confined flare, a novel phenomenon not reported
earlier.

\acknowledgments

We thank the SDO, RHESSI, and NoRH-NoRP teams for the high-quality EUV, X-ray, and microwave data. SDO is a mission of NASA's Living With a Star Program. We thank Dr. Jun Lin for constructive comments on the role of loop-loop dynamics, and Drs. Xin Cheng, and Wei Liu for helpful discussion on a preliminary analysis of the event. This work was supported by  NNSFC grants 41331068, 11790303, 11790304 (11790300), 11703017, and Natural Science Foundation of Shandong Province ZR2016AP13.


\begin{table}[!p]
  \centering
  \caption{RHESSI spectral fitting results for several intervals during Stage B} \label{Tab:1}
  \begin{tabular}{|c|p{1cm}|p{2.2cm}|p{1cm}|p{2.2cm}|p{1.8cm}|p{1.4cm}|p{1.0cm}|}
    \hline
    Time Interval (UT)&T1 (MK)&EM1 ($\times10^{49}$~cm$^{-3}$)&T2 (MK)&EM2 ($\times10^{49}$~cm$^{-3}$)&Break Energy~(keV)&Spectral Index&Chi-square \\
    \hline
    06:15:44-06:16:12&13.92&2.61&29.44&0.025&25.74&7.54&0.96\\
    \hline
    06:16:28-06:17:04&14.50&2.74&36.37&0.021&25.81&7.76&0.73\\
    \hline
    06:17:24-06:18:00&14.61&3.05&33.84&0.033&29.01&8.72&0.88\\
    \hline
    06:18:28-06:19:00&14.21&4.90&30.16&0.053&30.29&11.18&0.53\\
    \hline
    06:19:20-06:19:44&13.45&4.05&25.98&0.063&-&-&1.00\\
    \hline
  \end{tabular}
\end{table}

\end{document}